\documentclass[reprint,amsmath,amssymb,aps]{revtex4-2}

\usepackage{graphicx}
\usepackage{color}
\usepackage{dcolumn}
\usepackage{bm}
\usepackage{hyperref}

\usepackage{pifont}

\renewcommand{\eqref}[1]{(\ref{#1})}
\newcommand{\figref}[1]{Figure~\ref{#1}}
\newcommand{\appref}[1]{Appendix~\ref{#1}}

\newcommand{\phiT}{\phi_{\text{T}}}
\newcommand{\Var}{\text{Var}}

\begin{document}

\title{Self-assembly of biomolecular condensates with shared components}

\author{William M.~Jacobs}
\affiliation{Department of Chemistry, Princeton University, Princeton, NJ 08544, USA}

\date{\today}

\begin{abstract}
  Biomolecular condensates self-assemble when proteins and nucleic acids spontaneously demix to form droplets within the crowded intracellular milieu.
  This simple mechanism underlies the formation of a wide variety of membraneless compartments in living cells.
  To understand how multiple condensates with distinct compositions can self-assemble in such a heterogeneous system, we study a minimal model in which we can ``program'' the pairwise interactions among hundreds of species.
  We show that the number of distinct condensates that can be reliably assembled grows superlinearly with the number of species in the mixture when the condensates are allowed to share components.
  Furthermore, we show that we can predict the maximum number of distinct condensates in a mixture without knowing the details of the pairwise interactions.
  Simulations of condensate growth confirm these predictions and suggest that the physical rules governing the achievable complexity of condensate-mediated spatial organization are broadly applicable to biomolecular mixtures.
\end{abstract}

\maketitle

Many proteins and nucleic acids are spatially organized into biomolecular condensates within living cells~\cite{shin2017liquid}.
Condensates behave like liquid droplets, typically assembling via nucleation and growth while exhibiting a finite surface tension at the condensate--cytoplasm/nucleoplasm interface~\cite{berry2018physical}.
Importantly, the assembly of these structures is spontaneous, driven by net attractive interactions between disordered polypeptides~\cite{brangwynne2015polymer,choi2020physical} and by specific binding interactions among protein, DNA/RNA, and various client molecules~\cite{sanders2020competing,xing2020quantitative}.
Over the past decade, a large number of distinct types of condensates have been characterized.
Many of these examples assemble only under specific conditions and appear to be involved in a variety of key biological functions~\cite{shin2017liquid,banani2017biomolecular}, including stress response~\cite{sanders2020competing}, intracellular signaling~\cite{chong2016liquid}, and transcriptional control~\cite{sabari2020biomolecular}.
As such, it is important to understand how the self-assembly of a large number of condensates can be coordinated in time and space within a single intracellular compartment.

Previous efforts to describe spontaneous condensate assembly using simple theoretical models have provided insight into the typical behavior that can occur in biological mixtures with hundreds or thousands of components~\cite{sear2003instabilities,jacobs2017phase}.
However, these approaches have assumed a null model in which there is no inherent structure to the interactions among the various molecules in the mixture.
This is a poor assumption when considering biopolymer mixtures that have presumably evolved to assemble functional condensates.
For example, an intriguing situation is found in the nucleus, where putative transcriptional condensates self-assemble (or can be made to assemble~\cite{wei2020nucleated}) at specific target loci on the genome and share components while remaining mutually immiscible~\cite{sabari2020biomolecular,mir2018dynamic}.
If these structures form by a near-equilibrium mechanism akin to liquid--liquid phase separation, how many such structures can be simultaneously encoded by the intermolecular interactions and reliably assembled under spatiotemporal control?

\begin{figure}
  \includegraphics{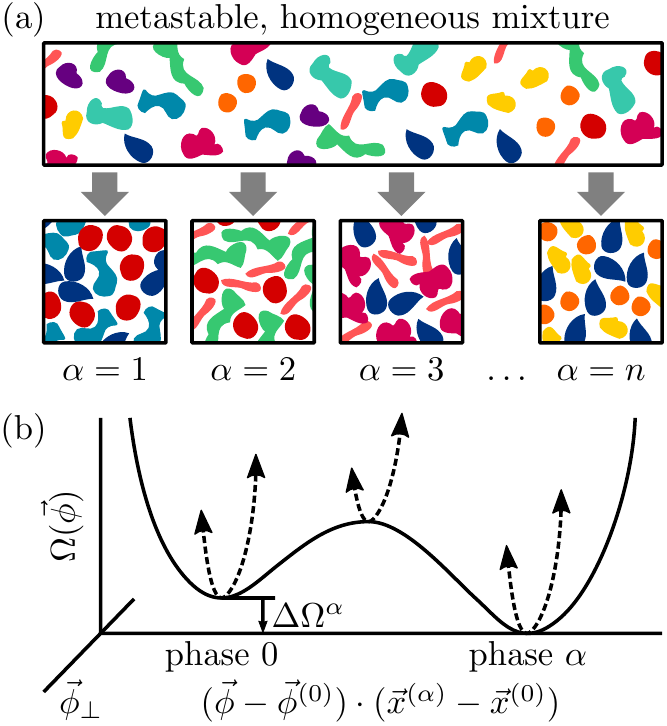}
  \caption{Encoding compositionally distinct droplets in a multicomponent mixture.
    (a)~We aim to encode $n$ target phases, each enriched in a specific subset of the components, that can be selectively assembled from a homogeneous mixture via nucleation and growth.
    (b)~To achieve this, each target phase must be a local minimum of the grand potential, $\Omega$.
    We sketch $\Omega$ along a linear path between the homogeneous and target-phase concentrations, $\vec\phi^{(0)}$ and $\vec\phi^{(\alpha)}$; orthogonal directions of concentration space, $\vec\phi_\perp$, are indicated by dashed curves.
    \label{fig:1}}
  \vskip-2.5ex
\end{figure}

To address this question, we consider the inverse problem of designing component-wise interactions to encode a set of target condensates.
In this way, we can determine the rules governing the assembly of condensates with shared components.
We specify this problem by proposing a set of $n$ target phases, each enriched in a specific subset of the $N$ species in the mixture (\figref{fig:1}a).
Defining the component volume fractions $\vec\phi$, the total volume fraction of all components ${\phiT \equiv \sum_{i=1}^N \phi_i}$, and the compositions ${x_i \equiv \phi_i / \phiT}$, we say that component $i$ is enriched in phase $\alpha$ if ${x_i^{(\alpha)} > x_i^{(0)}}$ and depleted otherwise, where the label 0 indicates the homogeneous phase.
We also specify the composition of the homogeneous phase, which has a total volume fraction of $\phiT^{(0)}$.
These choices establish the target phase behavior of the mixture.

In order for the target phases to self-assemble, each phase must be supersaturated with respect to the homogeneous phase and stable with respect to composition fluctuations.
These requirements imply that each phase is a local minimum of the dimensionless grand potential, ${\Omega(\vec\phi) \equiv F(\vec\phi) - \sum_i [\partial F(\vec\phi^{\,(0)})/ \partial \phi_i] \phi_i}$, where $F$ is the dimensionless Helmholtz free energy (\figref{fig:1}b).
The thermodynamic driving force for assembling each target phase, $\Delta\Omega^{(\alpha)}$, can be tuned by adjusting the volume fractions in the metastable mixture, $\vec\phi^{(0)}$.
Making the simplifying assumption that all interactions are pairwise (i.e., lacking any orientational dependence or multibody terms), we invoke the mean-field regular solution model,
\begin{equation*}
  F = \sum_{i=1}^N \phi_i \log \phi_i + (1 - \phiT)\log(1 - \phiT) - \sum_{i=1}^N\sum_{j=1}^N \chi_{ij} \phi_i \phi_j.
\end{equation*}\vskip-0.25ex
Our aim is thus to identify a symmetric component-wise interaction matrix, $\bm{\chi}$, that yields that target phase behavior.
Later, we shall show that our results translate to models with finite-range interactions.

Within the mean-field model, our inverse design problem requires that the following three conditions be met:
\begin{eqnarray}
  \label{eq:cond-depleted}
  x_i^{(\alpha)} \le \frac{\zeta}{NM_\alpha} \text{ if species }i \text{ is depleted}&\text{ in }&\text{phase }\alpha \\
  \label{eq:cond-mu}
  \sum_{i,j}\!\left[\phi_i^{(\alpha)}\!\phi_j^{(\alpha)}\!\!\! - \!2\delta_{ik}\! \left(\!\phi_j^{(\alpha)}\!\!\! - \!\phi_j^{(0)}\!\right)\!\! - \!\phi_i^{(0)}\!\phi_j^{(0)}\!\right]\!\chi_{ij} &=& -\!\log\!\frac{\phi_k^{(\alpha)}\!}{\phi_k^{(0)}\!}\quad\;\; \\
  \label{eq:cond-d2F}
  \left.\frac{\partial^2 F}{\partial \phi_i \partial \phi_j}\right|_\alpha = \frac{\delta_{ij}}{\phi_i^{(\alpha)}} + w(\bm{\chi}) - 2\chi_{ij} &\succ& 0,
\end{eqnarray}
where $M_\alpha$ is the number of components enriched in phase $\alpha$ and $w(\bm{\chi}) \equiv \exp[\sum_{i,j}\chi_{ij} (\phi_i^{(\alpha)}\!\phi_j^{(\alpha)} - \phi_i^{(0)}\!\phi_j^{(0)})] / (1 - \phiT^{(0)})$.
The first condition specifies that while we do not prescribe the precise compositions of the depleted components, they should nonetheless comprise a negligible volume fraction of a target phase.
The constant $\zeta$ must therefore be less than unity, as discussed below.
The second condition ensures that all phases are extrema of the grand potential.
We implicitly enforce $\Delta\Omega^{(\alpha)} \approx 0$ across all phases by making the approximation $\phiT^{(\alpha)} \simeq 1$; this approximation allows us to replace $\phi_i^{(\alpha)}$ with $x_i^{(\alpha)}$ everywhere and to neglect terms involving $x_i^{(\alpha)}$ from depleted components on the left-hand side of \eqref{eq:cond-mu}.
Finally, the third condition requires that $\partial^2 F/\partial\phi_i\partial\phi_j|_\alpha$ be positive definite to ensure the thermodynamic stability of the $\alpha$ phase.
Because $w(\bm{\chi})$ is positive and appears in all entries of this matrix, it negligibly affects the minimum eigenvalue.
Thus, for practical purposes, all three conditions are linear in $\bm{\chi}$.
Lastly, we assume for simplicity that the components do not interact with themselves, such that $\chi_{ii} = 0$, since such interactions equally stabilize all phases in which the $i$th component is enriched.

Even if a suitable interaction matrix exists for a set of target phases, it is typically not unique.
We therefore pick out the ``best'' solution, which maximizes the stability of the least stable phase, by minimizing the objective function
\begin{equation*}
  \mathcal{L}_{\text{SDP}} \equiv -\min_{\alpha=0,\ldots,n}\left\{\min_{i=1,\ldots,N}\left[\lambda_i^{(\alpha)}(\bm{\chi})\right]\right\} + \kappa\Var_{i<j}(\bm{\chi}),
\end{equation*}
where $\lambda_i^{(\alpha)}$ is the $i$th eigenvalue of $\partial^2 F/\partial \phi_i \partial \phi_j |_\alpha$ and $\Var_{i<j}$ is the variance of all elements in the upper triangle of $\bm{\chi}$.
The regularization parameter $\kappa > 0$ is introduced to ensure that the solution is unique.
Minimizing $\mathcal{L}_{\text{SDP}}$, subject to constraints \eqref{eq:cond-depleted} and \eqref{eq:cond-mu}, is a convex optimization problem that can be solved numerically using efficient semi-definite programming (SDP) algorithms~\cite{odonoghue2016conic,diamond2016cvxpy}.
Importantly, if ${\mathcal{L}_{\text{SDP}} < 0}$, then a solution exists to the mean-field inverse problem.

We can also devise a simpler optimization problem by approximating the minimum eigenvalue of $\partial^2 F$ using random matrix theory~\cite{wigner1967random}.
If the partition coefficient for depleted components, $\zeta/M_\alpha$, is sufficiently small, then we can approximate the SDP by the quadratic programming (QP) objective function
\begin{equation*}
  \mathcal{L}_{\text{QP}} \equiv \frac{\big(\phiT^{(0)}\big)^2}{N} \Var_{i<j}(\bm{\chi}) + \sum_{\alpha=1}^n\frac{1}{M_\alpha}\Var_{i<j}^{(\alpha)}(\bm{\chi}),
\end{equation*}
where $\Var_{i<j}^{(\alpha)}$ pertains to the subset of matrix elements for which both species $i$ and $j$ are enriched in phase $\alpha$, and the prefactors give equal weight to the stabilization of each phase in the random-matrix approximation~\cite{furedi1981eigenvalues}.

In addition to being easier to solve numerically, the QP formulation allows us to predict whether an interaction matrix exists for a specified set of target phases.
We consider a graph $G$ in which the $N$ components are vertices and draw an edge between a pair of components whenever they are both enriched in any one of the target phases; target phases are thus represented by cliques in $G$.
Within each target clique, minimizing $\Var_{i<j}^{(\alpha)}(\bm{\chi})$ with respect to \eqref{eq:cond-mu} has a unique solution.
However, in order to satisfy \eqref{eq:cond-depleted} and \eqref{eq:cond-mu} globally, there cannot be any cliques in $G$ with size greater than $M_\alpha$ that contain all $M_\alpha$ components of any target phase.
Such ``off-target'' cliques lead to the formation of chimeric phases, which combine components from multiple target phases.

We now arrive at a central conclusion of this work: With high probability, we can predict whether an interaction matrix exists for a set of target phases without knowing the target compositions.
For simplicity, we consider the case where all target phases are enriched in the same number of components, such that $M_\alpha = M$ for every target $\alpha$.
If the target compositions are uncorrelated, then we can treat the graph $G$ as an Erd\H{o}s--R\'{e}nyi random graph~\cite{frieze2016introduction}.
The probability of finding a clique of size $M+1$ that contains a target clique goes to zero as $N \rightarrow \infty$ if the probability of finding an edge between a pair of components obeys ${p_{\text{edge}} \lesssim (nN)^{-1/M}}$.
It follows that the number of target phases that can be reliably encoded scales as $n^* \sim N^{(2M-1)/(M+1)} \equiv N^\gamma$.
Therefore, we only need to know the number of components, the number of target phases, and the number of enriched components per target in order to predict whether a suitable interaction matrix is likely to exist.
This prediction provides a useful, albeit probabilistic, upper bound on the complexity of the phase behavior that can be achieved using pairwise interactions and shared components.

\begin{figure}
  \includegraphics{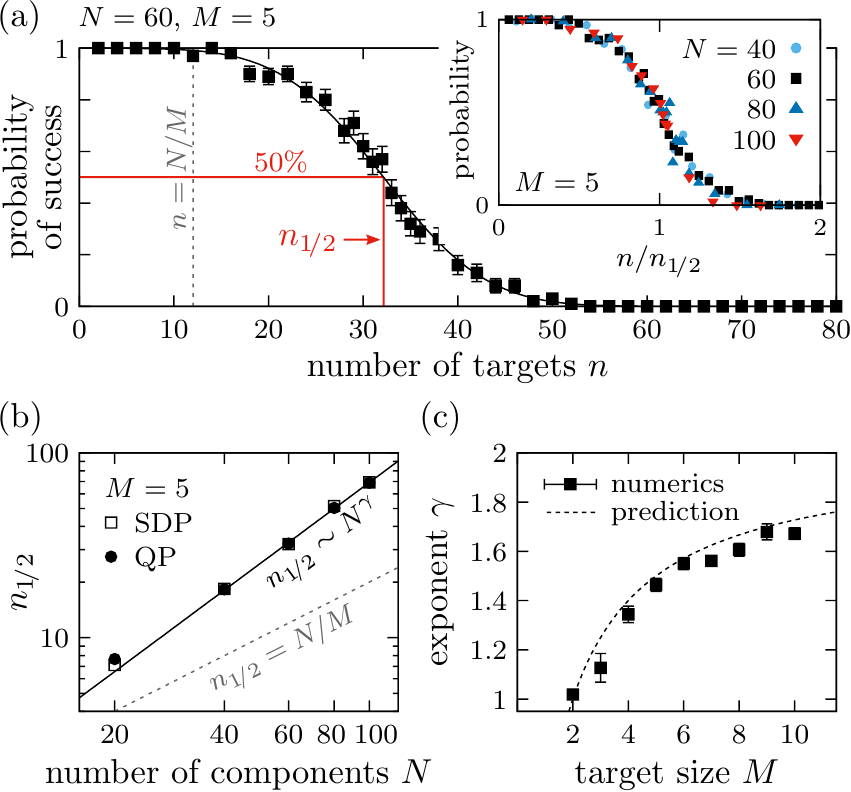}
  \vskip-0.25ex
  \caption{Mean-field predictions of optimal target encoding.
    (a)~The probability that a set of $n$ randomly generated targets can be stably encoded in the mean-field model.
    \textit{Inset}: The probability of successful encoding roughly follows a master curve when scaled by $n_{1\!/2}$, the number of targets that can be encoded with 50\% probability.
    (b)~For ${N \gg M}$, numerical calculations of $n_{1\!/2}$ obtained via quadratic programming (QP) agree well with those obtained by semi-definite programming (SDP), showing that $n_{1\!/2}$ follows a power law as a function of the number of components.
    (c)~The scaling exponent $\gamma$ increases with the number of species enriched in each target, $M$, as predicted by graph-theoretical arguments (see text).
    \label{fig:2}}
  \vskip-1.5ex
\end{figure}

We test our predictions by generating random sets of target phases, each enriched in $M$ components, and numerically optimizing the interaction matrix $\bm{\chi}$ using both the SDP and QP formulations.
We then numerically check whether the optimal $\bm{\chi}$ matrix in fact results in the desired grand potential landscape as depicted in \figref{fig:1}b.
To this end, we minimize $\Omega(\vec\phi)$ within the basin of attraction of each target phase and verify that the local minimum is consistent with the target composition.
Representative results of these calculations are shown in \figref{fig:2}a, where by varying the number of encoded targets, we empirically determine the probability that a $\bm{\chi}$ matrix exists for a random set of target phases.
We first observe that the number of targets that can be encoded is typically much larger than $N/M$, the maximum number of phases possible if every species is enriched in at most one target.
Second, we find evidence of the predicted thresholding transition: The probability of successful encoding drops rapidly from near unity to near zero in a manner that depends on both $N$ and $M$.
We characterize this transition by $n_{1/2}$, the number of targets for which the probability of successful encoding is 50\%.
Rescaling $n$ by $n_{1/2}$ reveals that the data collapse well across mixtures with different numbers of components (\figref{fig:2}a,\textit{inset}).

We next test the graph-theoretical scaling prediction for the thresholding transition.
Plotting $n_{1/2}$ versus $N$, we confirm that the threshold indeed follows a power law once in the scaling regime, $N \gg M$ (\figref{fig:2}b).
We also verify that the thresholding transition occurs at the same number of targets regardless of whether the $\bm{\chi}$ matrix is determined via SDP and QP optimization.
This is the case due to our choice of ${\zeta = 10^{-3}}$; if we choose $\zeta$ closer to unity, implying that depleted components are not as strongly repelled from the target phases, then $n_{1/2}$ increases for the SDP solutions, although the power law scaling remaining unchanged (data not shown).
Finally, we compare the empirical scaling exponents, determined from power law fits to the thresholding data, to the graph-theoretical predictions (\figref{fig:2}c).
In close agreement with the predictions, we find that the number of targets that can be reliably encoded scales superlinearly with $N$ whenever three or more components are enriched in each target.
This is a second key conclusion: Because the number of phases that can be reliably programmed may grow faster than the number of components, multicomponent mixtures can support dramatically more complex phase behavior than simple model systems that contain only a handful of components.

\begin{figure}
  \includegraphics{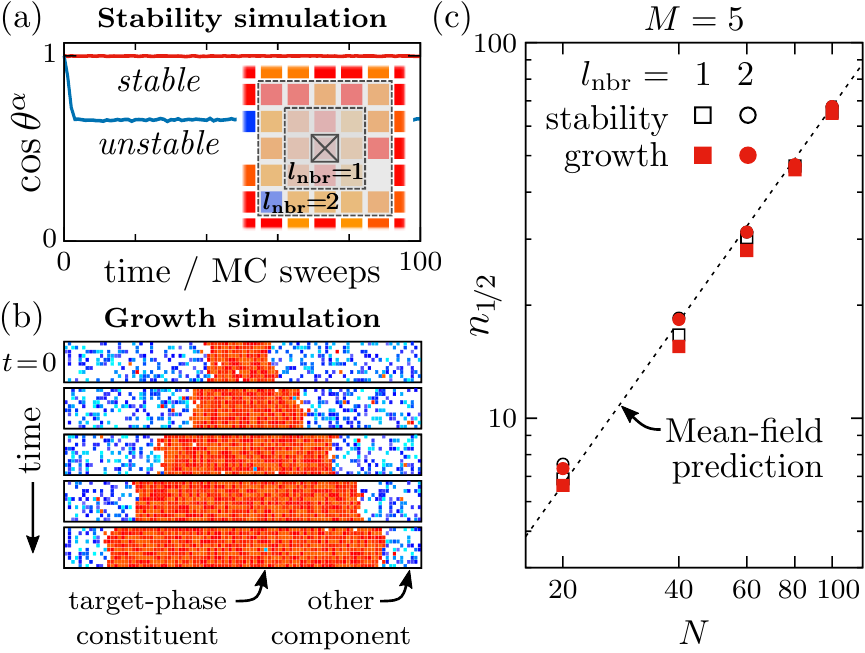}
  \caption{Lattice simulations with finite-range interactions follow the mean-field predictions.
    (a)~Bulk phases are considered stable if the composition of the lattice fluctuates near the target composition, such that $\cos\theta^{(\alpha)} \simeq 1$.
    \textit{Inset}: Illustration of first- and second-nearest neighbor shells on the three-dimensional lattice.
    (b)~Growth from a metastable gas phase is considered successful if the composition of the droplet matches that of the initial seed at time ${t = 0}$. Components colored red are enriched in the target phase, while blue components are depleted.
    (c)~The values of $n_{1\!/2}$ obtained by simulating stability and growth approach the mean-field predictions when $N \gg M$.
    However, short-range interactions (i.e., $l_{\text{nbr}}=1$) slightly reduce the number of targets than can be stably encoded and then assembled via droplet growth.
    \label{fig:3}}
  \vskip-1ex
\end{figure}

Having analyzed the programmability of target phases in the mean-field model, we now ask whether these results translate to a model of a fluid with finite-range interactions.
For this purpose, we study a three-dimensional lattice gas model with Hamiltonian
\begin{equation*}
  \mathcal{H}_{\text{LG}} = \sum_{\langle r,s \rangle_{\text{nbr}}} \sum_{i,j} \epsilon_{ij} c_{ir} c_{js} - \mu \sum_r \sum_i c_{ir},
\end{equation*}
where $r$ and $s$ indicate lattice sites, $\bm{\epsilon}$ is the interaction matrix, ${c_{ir} = 1}$ if lattice site $r$ is occupied by a particle of type $i$ and is zero otherwise, and $\mu$ is the chemical potential of each particle.
The first sum indicates that pairs of particles interact within a distance of $l_{\text{nbr}}$ on the lattice (\figref{fig:3}a,\textit{inset}).
Because the analogues of constraints \eqref{eq:cond-mu} and \eqref{eq:cond-d2F} are nonlinear in this model, we simply apply the mean-field QP solution, $\bm{\chi}$, and rescale it such that the net interaction energy within each target phase, ${\epsilon_{\text{target}} \equiv \sum_{i,j} \epsilon_{ij} (x_i^{(\alpha)}\!x_j^{(\alpha)}\! - x_i^{(0)}\!x_j^{(0)})}$, is below the critical point.
We then test whether the designed interaction matrix $\bm{\epsilon}$ codes for the self-assembly of the target phases by performing two types of grand-canonical Monte Carlo simulations~\cite{frenkel2001understanding}.
First, we perform ``stability'' simulations initialized from each of the target phases, as well as from the metastable, homogeneous phase (\figref{fig:3}a).
By computing the similarity between the target volume fractions and the volume fractions $\vec\phi$ observed in the simulation, ${\cos \theta^{(\alpha)} \equiv (\vec\phi^{(\alpha)} \cdot \vec\phi) / |\vec\phi^{(\alpha)}| |\vec\phi|}$, we can determine whether the target phase is stable with respect to composition fluctuations.
Second, we perform ``growth'' simulations in a slab geometry, starting from a post-critical nucleus of the target phase amidst a metastable gas of the homogeneous phase (\figref{fig:3}b).
We classify such a simulation as successful if a phase with the target composition grows to fill the simulation box.
As noted below, the outcome of these simulations depends on both $\mu$ and $\epsilon_{\text{target}}$.

Our third key conclusion is that the scaling behavior predicted by the mean-field model holds for both the stability and growth simulations.
A representative summary of these calculations is shown in \figref{fig:3}c.
Although the empirical value of $n_{1/2}$ for each type of simulation is generally lower than that of the mean-field model, we observe that these values converge as the number of components increases.
In the cases where $\bm{\chi}$ solves the mean-field inverse problem but the corresponding $\bm{\epsilon}$ fails in the growth simulations, the cause is typically either a lower free-energy barrier between phases or the nucleation of off-target phases at the droplet--gas phase interface in the lattice model.
We note that increasing the interaction range from $l_{\text{nbr}}=1$ to $l_{\text{nbr}}=2$, thereby making the lattice model more ``mean-field-like,'' consistently increases the number of targets that can be reliably assembled.

\begin{figure}
  \includegraphics{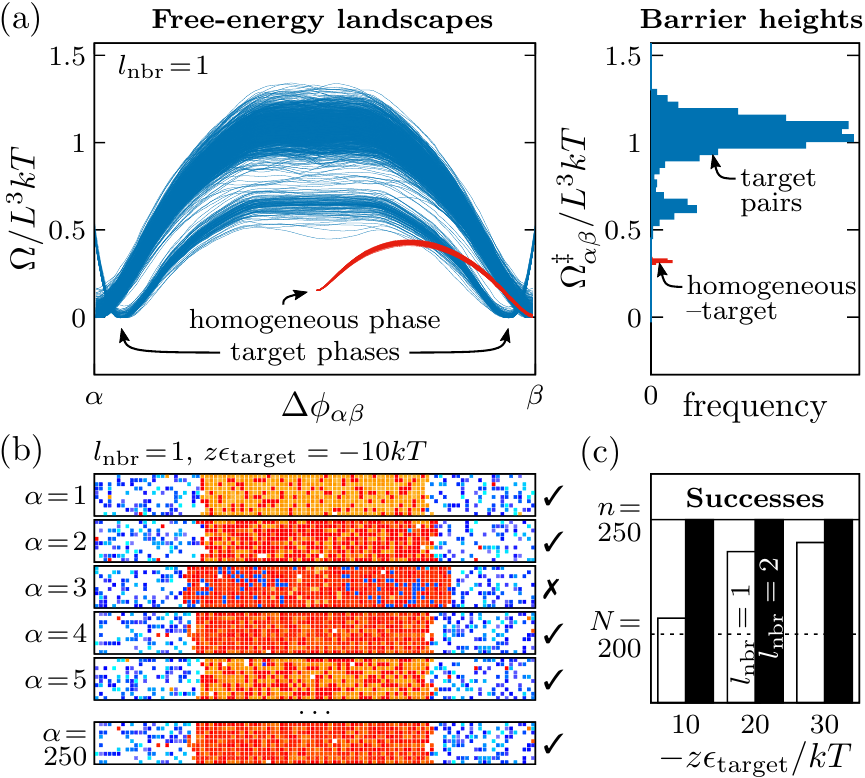}
  \caption{Encoding 250 compositionally distinct target phases, each enriched in five species, using only 200 components.
    (a,\textit{left})~Projections of the grand potential on an $L \times L \times L$ lattice with $L=6$ and $l_{\text{nbr}}=1$.
    Each curve shows the free energy surface along a linear path in concentration space, parameterized by $\Delta\phi_{\alpha\beta}$, between a pair of target phases (blue) or between the homogeneous phase and a target phase (red).
    (a,\textit{right})~The corresponding distribution of barrier heights---i.e., saddle points on the grand potential landscape---between pairs of phases, $\Omega^\ddagger_{\alpha\beta}$.
    (b)~Snapshots from $l_{\text{nbr}}=1$ growth simulations, seeded one-at-a-time by each of the 250 target phases.
    Species belonging to the target phase are colored red/yellow, while those depleted in the target phase are colored blue/cyan.
    (c)~The number of successfully grown target phases as a function of the mean interaction energy within each target phase, $\epsilon_{\text{target}}$, and the interaction range $l_{\text{nbr}}$; $z=6$ is the lattice coordination number.
    \label{fig:4}}
  \vskip-1ex
\end{figure}

We conclude by demonstrating the possibility of encoding and self-assembling a set of compositionally distinct condensates that outnumber the components in a mixture, as suggested by the superlinear scaling of $n_{1/2}$ in \figref{fig:2}c.
As an example, we apply the QP formulation in an attempt to encode 250 targets, each enriched in five species, using only 200 components.
We use Wang--Landau simulations~\cite{wang2001efficient,jacobs2013predicting} to confirm that these target phases correspond to local minima of the grand potential and are supersaturated relative to the homogeneous mixture in the $l_{\text{nbr}}=1$ lattice model (\figref{fig:4}a,\textit{left}).
From these calculations, we are able to estimate the barrier heights, which are proportional to the macroscopic surface tensions, between pairs of target phases and between target phases and the homogeneous phase (\figref{fig:4}a,\textit{right}).
We find that growth simulations occasionally fail to reproduce the target composition from a post-critical seed when $l_{\text{nbr}}=1$ (\figref{fig:4}b).
However, the fraction of successfully assembled targets increases with both the magnitude of the interaction strength and the number of neighbors with which particles interact, as both of these strategies increase the barrier heights between the target phases (\figref{fig:4}c); on a practical note, the latter requirement of a large effective coordination number is likely to be satisfied in condensed phases of disordered biopolymers~\cite{colby2003polymer}.
Thus, by introducing one or more seeds of each target phase into a single closed system, our simulations demonstrate the possibility of \textit{simultaneously} self-assembling a greater number of condensates than there are components using pairwise interactions alone.
While this result might appear at odds with the Gibbs Phase Rule, the paradox is resolved by noting that condensate self-assembly can only be sustained by a finite thermodynamic driving force and is thus an inherently non-equilibrium process.
Ultimately, some condensates must grow at the expense of others as the system relaxes to equilibrium unless coarsening is arrested by another non-equilibrium mechanism~\cite{weber2019physics}.

In summary, we have shown that the number of condensates that can be encoded in a multicomponent mixture and reliably self-assembled can increase superlinearly with the number of species when condensates share components.
The probability that an encoding exists for a random set of target compositions undergoes a thresholding transition, implying that the capacity of a mixture to self-assemble the target condensates can be predicted by knowing only the number of components in the mixture, the number of species enriched in each target, and the number of targets.
These predictions apply to any molecular system in which pairwise interactions among species can be programmed, either synthetically or through an evolutionary process.
Furthermore, our results highlight the vast possibilities of programmable self-assembly in systems with thousands of components, as is the case in many biological mixtures, even in the absence of orientationally specific saturating bonds~\cite{murugan2015multifarious}.
An intriguing possibility is that condensate assembly might play a role in intracellular information processing, whereby the combinatorial complexity described here is exploited to create spontaneous spatial organization in response to stimuli, which take the form of heterogeneous nucleation sites.
This model may thus have implications for understanding spatiotemporal control of transcription via condensate assembly in the nucleus.

This work was carried out with financial support from Princeton University and computational resources provided by Princeton University Research Computing.

%

\clearpage
\onecolumngrid
\appendix
\section*{Supplementary Information}

\section{Specifying the inverse problem}
\label{sec:inverse-problem}

We define the target phase behavior by specifying the homogeneous phase volume fractions $\vec\phi^{(0)}$, as well as the composition, $x_i^{(\alpha)}$, for each enriched component $i$ in each target phase $\alpha$.
The compositions of the depleted components in each target phase are not specified; however, we require that all the depleted components comprise a volume fraction within the target phase that is less than $1/M_\alpha$, where $M_\alpha$ is the number of enriched components in the $\alpha$ phase.
Therefore, we arrive at the first constraint in the main text,
\begin{equation}
  \label{eq:constraint-1}
  x_i^{(\alpha)} \le \frac{\zeta}{NM_\alpha}\text{ if species }i\text{ is depleted in phase }\alpha,
\end{equation}
where $\zeta < 1$ is a constant that we specify when solving the convex optimization problem (see \appref{sec:convex-optimization}).

The second constraint in the main text derives from the conditions for phase coexistence in the mean-field model.
Phase coexistence requires that the dimensionless chemical potentials of all components,
\begin{equation}
  \mu_i = \frac{\partial F}{\partial \phi_i} = \log \phi_i - \log(1 - \phiT) - 2\sum_j \chi_{ij} \phi_j,
\end{equation}
and the dimensionless osmotic pressure,
\begin{equation}
  \Pi = \sum_i \phi_i \mu_i - F = -\log(1 - \phiT) - \sum_{i,j} \chi_{ij} \phi_i \phi_j,
\end{equation}
be equal across all phases.
We set $\mu_i^{(\alpha)} = \mu_i^{(0)}$ and $\Pi^{(\alpha)} = \Pi^{(0)}$, and then eliminate the terms involving $\phiT^{(\alpha)}$ from the equations for the chemical potentials.
Approximating $\phiT^{(\alpha)} \approx 1$ for each condensed target phase allows us to write the second constraint in the main text as
\begin{equation}
  \label{eq:constraint-2}
  \sum_{i,j}\left[x_i^{(\alpha)} x_j^{(\alpha)} \bm{1}_i^{(\alpha)} \bm{1}_j^{(\alpha)}\! - 2\delta_{ik} (x_j^{(\alpha)} \bm{1}_j^{(\alpha)}\! - \phi_j^{(0)}) - \phi_i^{(0)}\phi_j^{(0)}\right]\chi_{ij} = -\log\frac{x_k^{(\alpha)}}{\phi_k^{(0)}},
\end{equation}
where ${\bm{1}_i^{(\alpha)} = 1}$ if species $i$ is enriched in phase $\alpha$ and is zero otherwise; as a result of this approximation, the osmotic pressures of the various phases are not exactly equal, and we obtain a grand potential landscape in which the condensed phases are supersaturated relative to the homogeneous phase.
We note that the supersaturation can be adjusted by tuning the volume fractions in the homogeneous phase, $\vec\phi^{(0)}$.

We obtain the third constraint in the main text by eliminating the term $1/(1-\phiT^{(\alpha)})$ from $\partial^2 F/\partial \phi_i \partial \phi_j |_\alpha$ by again using the approximation $\Pi^{(\alpha)} \simeq \Pi^{(0)}$:
\begin{equation}
  \label{eq:constraint-3}
  \left.\frac{\partial^2 F}{\partial \phi_i \partial \phi_j}\right|_\alpha = \frac{\delta_{ij}}{\phi_i^{(\alpha)}} + w(\bm{\chi}) - 2\chi_{ij} \succ 0.
\end{equation}
As noted in the main text, $w(\bm{\chi})$ has a negligible effect on the minimum eigenvalue of $\partial^2 F/\partial \phi_i \partial \phi_j |_\alpha$.
We therefore approximate $w(\bm{\chi})$ by a constant expression when solving the optimization problem,
\begin{equation}
  w(\bm{\chi}) \simeq \frac{\exp\left[\langle\chi\rangle_{\text{target}} - \big(\phiT^{(0)}\big)^2 \langle\chi\rangle_{\text{homo}}\right]}{1 - \phiT^{(0)}},
\end{equation}
where, assuming the enriched components comprise equal volume fractions in all target phases, we have defined the expected averages of $\chi$ in the target, ${\langle\chi\rangle_{\text{target}} \equiv [\log(N / M \phiT^{(0)}) + (2\phiT^{(0)} - (\phiT^{(0)})^2)(N-M)\log(\zeta/N) / 2N] / (1 - \phiT^{(0)})^2}$, and homogeneous phases, ${\langle\chi\rangle_{\text{homo}} \equiv \langle\chi\rangle_{\text{target}} + (N - M)\log(\zeta/N) / 2N}$.
With these approximations, the inverse problem constraints \eqref{eq:constraint-1}, \eqref{eq:constraint-2}, and \eqref{eq:constraint-3} are linear in $\bm{\chi}$ and only involve quantities specified in the problem definition, i.e., $\vec\phi^{(0)}$ and the compositions $\{x_i^{(\alpha)}\}$ of the enriched components.

\section{Solving the convex optimization problems}
\label{sec:convex-optimization}

When performing numerical tests using the mean-field model, we generate each target phase by randomly choosing $M$ of the $N$ species.
We repeat this procedure to determine which species are enriched in each of the $n$ target phases, and we verify that the targets generated in this way are in fact unique.
The calculations presented in the main text assume an equimolar homogeneous phase with $\phiT^{(0)} = 0.25$.

When searching for the optimal $\bm{\chi}$ matrix within the SDP and QP formulations, we seek to minimize the objective functions
\begin{equation}
  \label{eq:LSDP}
  \mathcal{L}_{\text{SDP}} \equiv -\min_{\alpha=0,\ldots,n}\left\{\min_{i=1,\ldots,N}\left[\lambda_i^{(\alpha)}(\bm{\chi})\right]\right\} + \kappa\Var_{i<j}(\bm{\chi})
\end{equation}
or
\begin{equation}
  \label{eq:LQP}
  \mathcal{L}_{\text{QP}} \equiv \frac{(\phiT^{(0)})^2}{N} \Var_{i<j}(\bm{\chi}) + \sum_{\alpha=1}^n\frac{1}{M}\Var_{i<j}^{(\alpha)}(\bm{\chi})
\end{equation}
subject to the constraints
\begin{eqnarray}
  \sum_{i,j}\left[x_i^{(\alpha)}\!x_j^{(\alpha)} \bm{1}_i^{(\alpha)} \bm{1}_j^{(\alpha)}\! - 2\delta_{ik} (x_j^{(\alpha)} \bm{1}_j^{(\alpha)}\! - \phi_j^{(0)}) - \phi_i^{(0)}\!\phi_j^{(0)}\right]\chi_{ij} &=& -\log\frac{x_k^{(\alpha)}}{\phi_k^{(0)}} \;\;\text{if species }k\text{ is enriched in phase }\alpha, \\
  \sum_{i,j}\left[x_i^{(\alpha)}\!x_j^{(\alpha)} \bm{1}_i^{(\alpha)} \bm{1}_j^{(\alpha)}\! - 2\delta_{ik} (x_j^{(\alpha)} \bm{1}_j^{(\alpha)}\! - \phi_j^{(0)}) - \phi_i^{(0)}\!\phi_j^{(0)}\right]\chi_{ij} &\ge& -\log\frac{\zeta}{NM\phi_k^{(0)}} \;\;\text{if species }k\text{ is depleted in phase }\alpha.\qquad
\end{eqnarray}
We obtain the QP objective function by writing the stability matrix, $\partial^2 F/\partial\phi_i\partial\phi_j$, in the form
\begin{equation}
  \left.\frac{\partial^2 F}{\partial \phi_i \partial \phi_j}\right|_\alpha =
  \left[\begin{array}{@{}c|c@{}}
    A & B^\top \\
    \hline
    B & C^{\phantom\top} \end{array}\right],
\end{equation}
where the submatrix $A$ represents the $M \times M$ block of enriched components in the $\alpha$ phase.
The stability condition \eqref{eq:constraint-3} is equivalent to ${(\partial^2F) / C \equiv A - B^\top C^{-1} B \succ 0}$.
Furthermore, because the submatrix $C$ pertains to the components that are depleted in the $\alpha$ phase, the inverse of $C$ has eigenvalues clustered near $x_i^{(\alpha)} \phiT^{(\alpha)} \lesssim \zeta / N M_\alpha$.
Finally, by treating $(\partial^2 F)/C$ as a symmetric random matrix, we can seek to maximize its lowest eigenvalue by minimizing the variance of its off-diagonal elements~\cite{wigner1967random,furedi1981eigenvalues}.
This random-matrix approximation is a reasonable assumption if the target-phase compositions are uncorrelated.
The prefactors in \eqref{eq:LQP} are determined from the scaling behavior of the lowest eigenvalue, $\lambda_{\text{min}}$, of $\partial^2 F/\partial\phi_i\partial\phi_j$, assuming that this matrix can be treated as a symmetric random matrix~\cite{furedi1981eigenvalues}:
\begin{equation}
  \lambda_{\text{min}}^{(0)} \simeq \frac{N}{\phiT^{(0)}} - 2\sqrt{N\Var_{i<j}(\bm{\chi})}
  \qquad\text{and}\qquad
  \lambda_{\text{min}}^{(\alpha)} \simeq M - 2\sqrt{M\Var_{i<j}^{(\alpha)}(\bm{\chi})}\;\text{ for }\alpha=1,\ldots,n.
\end{equation}
Thus, if $\zeta$ is sufficiently small, the mean-field inverse problem has a solution if ${\mathcal{L}_{\text{QP}} \lesssim (n+1)/4}$, in which case the $\bm{\chi}$ matrix that minimizes $\mathcal{L}_{\text{QP}}$ is a good approximation of the $\bm{\chi}$ matrix that minimizes $\mathcal{L}_{\text{SDP}}$.

In the results presented in the main text, the depleted-component partition-coefficient constant $\zeta$ and the regularization constant $\kappa$ are chosen to be $10^{-3}$ and $10^{-5}\!$, respectively.
Numerical optimization is performed using the SCS solver~\cite{odonoghue2016conic} within the CVXPY framework~\cite{diamond2016cvxpy}.

\section{Predicting the thresholding transition}
\label{sec:thresholding}

We now elaborate on the random graph arguments discussed in the main text.
The probability that a pair of components $i$ and $j$ are both enriched in any one of the target phases is
\begin{equation}
  p_{\text{edge}} = 1 - \left[1 - \frac{M(M - 1)}{N(N-1)}\right]^n \simeq \frac{nM^2}{N^2}.
\end{equation}
The expectation value for the number of $(M+1)$-sized cliques that contain an $M$-sized target phase in the random graph is
\begin{equation}
  \label{eq:EK}
  \mathbb{E} \le n (N-M) p_{\text{edge}}^{M}.
\end{equation}
With high probability, a random graph does not contain such a clique if $\mathbb{E} < 1$; thus, we obtain a threshold for $n$,
\begin{equation}
  \label{eq:n_threshold}
  n^* \sim N^{(2M-1)/(M+1)},
\end{equation}
when $N \gg M$.
As discussed in the main text, we expect that $n_{1/2}$ will scale according to \eqref{eq:n_threshold} in the mean-field model.
We note that the threshold for the appearance of a random $(M+1)$-sized clique in an Erd\H{o}s--R\'{e}nyi random graph $G(N,p_{\text{edge}})$ with uncorrelated edges scales as~\cite{frieze2016introduction}
\begin{equation}
  n^*_{K_{M+1}} \sim N^{2(M-1)/M}.
\end{equation}
Our predicted scaling for $n_{1/2}$ grows more slowly with respect to $N$ than $n^*_{K_{M+1}}$ when $N \gg M$ because \eqref{eq:EK} accounts for the fact that $M$-sized cliques exist in the graph by design.
Thus, we expect to encounter $(M+1)$-sized cliques that directly destabilize target phases before we encounter completely random cliques of size $M+1$ or greater.
This expectation is borne out by the close agreement between our predicted scaling, \eqref{eq:n_threshold}, and the empirical scaling exponents determined from power law fits to the data in \figref{fig:2}b,c in the main text.

\section{Performing stability and growth simulations}
\label{sec:simulations}

We perform grand-canonical Monte Carlo (GCMC) simulations of stability and growth using a three-dimensional cubic lattice model.
As described in the main text, we obtain the interaction matrix, $\bm{\epsilon}$, from the mean-field solution by rescaling $\bm{\chi}$.
We then set the chemical potential, $\mu$, to be the same for all components; $\mu$ is chosen such that the target condensed phases are supersaturated relative to the homogeneous phase, and the chemical potential of a vacancy is zero.
Simulations are then carried out by proposing particle exchanges at every lattice site with equal probability and accepting these moves according to the standard Metropolis criterion~\cite{frenkel2001understanding}.

When running stability simulations, we initialize a $10 \times 10 \times 10$ cubic lattice with periodic boundary conditions in each one of the target phases, including the homogeneous phase.
This is accomplished by randomly assigning particles to lattice sites such that the lattice composition is consistent with the target composition, $\vec x^{(\alpha)}$.
We then evolve the lattice via GCMC simulation, keeping track of the order parameter $\cos\theta^{(\alpha)}$.
By construction, $\cos\theta^{(\alpha)} = 1$ at the start of each simulation.
However, $\cos\theta^{(\alpha)}$ is found to decrease markedly whenever the initial phase is unstable with respect to composition fluctuations; we define $\cos\theta^{(\alpha)} \ge 0.95$ to be the threshold for a ``successful'' stability simulation.
By running these simulations for only 100 Monte Carlo sweeps, we ascertain whether the free-energy barriers surrounding each one of the target phases are all higher than a few $kT$.
We note that when target phases are found to be unstable, the simulation usually ends up in a condensed-phase free-energy basin that is enriched in more than $M$ components.
Such chimeric phases tend to have greater compositional entropy and are thus lower in free energy than any of the $M$-component target phases.

When running growth simulations, we initialize a $100 \times 10 \times 10$ periodic cubic lattice with a $10 \times 10 \times 10$ seed of the target phase surrounded by the metastable, homogeneous gas phase.
The initial configurations of both the seed and the gas phase are prepared in the same manner as the stability simulations described above.
When sufficiently supersaturated relative to the gas phase, the seed grows by advancing both interfaces with the gas phase at a constant velocity parallel to the long axis of the simulation box.
Simulations are halted once the condensed phase occupies at least 90\% of the simulation box.
At this point, we classify a simulation as ``successful'' if the composition of the lattice occupied by the condensed phase matches the composition of the target phase using the same order parameter and threshold as above, i.e., if $\cos\theta^{(\alpha)} \ge 0.95$.

\section{Calculating free-energy landscapes}
\label{sec:landscapes}

We perform Wang--Landau simulations~\cite{wang2001efficient} to calculate the grand-potential free-energy surfaces between pairs of target phases in the lattice model, as shown in \figref{fig:4}a in the main text.
Following the method introduced in~\cite{jacobs2013predicting}, we define an order parameter $\Delta\phi_{\alpha\beta}$ for a pair of phases $\alpha$ and $\beta$,
\begin{equation}
  \Delta\phi_{\alpha\beta}(\vec\phi) \equiv (\vec\phi - \vec\xi_{\alpha\beta}) \cdot \hat\nu_{\alpha\beta},
\end{equation}
where ${\vec\xi_{\alpha\beta} \equiv (\vec\phi^{(\alpha)} + \vec\phi^{(\beta)}) / 2}$ and ${\hat\nu_{\alpha\beta} \equiv (\vec\phi^{(\beta)} - \vec\phi^{(\alpha)}) / |\vec\phi^{(\beta)} - \vec\phi^{(\alpha)}|}$.
To constrain the simulation to explore the phase space near phases $\alpha$ and $\beta$, we add a harmonic potential in directions of concentration space orthogonal to $\vec\xi_{\alpha\beta}$,
\begin{equation}
  U_{\alpha\beta}(\vec\phi) \equiv k_\perp \big|(\vec\phi - \vec\xi_{\alpha\beta}) - [(\vec\phi - \vec\xi_{\alpha\beta}) \cdot \hat\nu_{\alpha\beta}] \hat\nu_{\alpha\beta}\big|^2.
\end{equation}
To improve the efficiency of the simulation, particle exchanges from a lattice site occupied by a particle (or a vacancy) of type $i$ to a particle (or a vacancy) of type $j$ are proposed with probability
\begin{equation}
  p_{\text{gen}}(i \rightarrow j) = \begin{cases}
    0.5 &\text{if }j\text{ is a vacancy} \\
    (0.5 - 0.01) / M_\beta &\text{if }j\text{ is enriched in phase }\beta \\
    0.01 / (N - M_\beta) &\text{if }j\text{ is depleted in phase }\beta
  \end{cases}
\end{equation}
if $\alpha$ is the homogeneous phase and $\beta$ is a condensed target phase, or with probability
\begin{equation}
  p_{\text{gen}}(i \rightarrow j) = \begin{cases}
    (1 - 0.01) / M_{\alpha\beta} &\text{if }j\text{ is enriched either in phase }\alpha\text{ or in phase }\beta \\
    0.01 / (N + 1 - M_{\alpha\beta}) &\text{otherwise, including if }j\text{ is a vacancy},
  \end{cases}
\end{equation}
where $M_{\alpha\beta}$ is the number of components that are enriched either in phase $\alpha$ or in phase $\beta$, if $\alpha$ and $\beta$ are both condensed target phases.
The Metropolis acceptance criterion is appropriately modified to account for these non-uniform generation probabilities.

The free-energy landscapes presented in the main text are obtained using the combined potential, ${\mathcal{H}_{\text{LG}} + U_{\alpha\beta}}$, on an $L \times L \times L$ periodic lattice with ${L = 6}$ and ${k_\perp = 10^4}$.
\figref{fig:4}a shows landscapes for 1,000 randomly selected pairs of target phases and for all 250 homogeneous-target phase pairs.
The barrier heights, $\Omega^\ddagger$, are computed by finding the maximum on the computed free-energy surface between phases $\alpha$ and $\beta$ at coexistence between the two phases.

\end{document}